\documentclass[10pt]{article}

\usepackage{amsmath}
\usepackage{amssymb}

\usepackage{graphicx}

\usepackage{cite}

\usepackage{color} 

\usepackage{setspace} 
\usepackage[labelfont=bf,labelsep=period,justification=raggedright]{caption}

\makeatletter
\renewcommand{\@biblabel}[1]{\quad#1.}
\makeatother

\date{}

\begin{document}

\begin{flushleft}
{\Large
\textbf{FragIt: A Tool to Prepare Input Files for Fragment Based Quantum Chemical Calculations}
}
\\
Casper Steinmann$\ast$, 
Mikael W Ibsen,
Anne S Hansen,
Jan H Jensen
\\
Department of Chemistry, University of Copenhagen, Copenhagen, Denmark
\\
$\ast$ corresponding author, E-mail: steinmann@chem.ku.dk
\end{flushleft}

\section*{Abstract}
Near linear scaling fragment based quantum chemical calculations are becoming increasingly popular for treating large systems with high accuracy and is an active field of research. However, it remains difficult to set up these calculations without expert knowledge. To facilitate the use of such methods, software tools need to be available to support these methods and help to set up reasonable input files which will lower the barrier of entry for usage by non-experts. Previous tools relies on specific annotations in structure files for automatic and successful fragmentation such as residues in PDB files. We present a \emph{general} fragmentation methodology and accompanying tools called FragIt to help setup these calculations. FragIt uses the SMARTS language to locate chemically appropriate fragments in large structures and is applicable to fragmentation of any molecular system given suitable SMARTS patterns. We present SMARTS patterns of fragmentation for proteins, DNA and polysaccharides, specifically for D-galactopyranose for use in cyclodextrins. FragIt is used to prepare input files for the Fragment Molecular Orbital method in the GAMESS program package, but can be extended to other computational methods easily.

\section*{Introduction}
The need to compute molecular properties for larger and
larger systems with desirable accuracy has led to the development of novel
methods such as fragmentation methods\cite{gordon2012fragmentation}. In fragmentation methods, a large system
is divided into several smaller subsystems called fragments.  Each fragment is
treated with some \emph{ab initio} level of theory and different
methods\cite{gao1997energy,xie2007design,suarez2009thermochemical,fedorov2007extending,
fedorov2009fragment,mayhall2011molecules} include the surrounding environment in different ways.

In this work, we are interested in setting up Fragment Molecular Orbital (FMO)\cite{fedorov2007extending,fedorov2009fragment} and Effective Fragment Molecular Orbital (EFMO)\cite{steinmann2010effective,steinmann2012effective} calculations, but our method is extensible to other fragment based methods. In the FMO method, each fragment is polarized by the presence of the Coulomb field of all other fragments. The underlying equations allow for a systematic improvement of the energy by considering pairs and optionally triples of fragments\cite{fedorov2006three}, the latter often within milihartree accuracy of the corresponding \emph{ab initio} energy. FMO supports correlated treatment of one or more fragments\cite{fedorov2004second,fedorov2005coupled,fedorov2007accuracy} as well the possibility of obtain excitation energies with good accuracy.\cite{chiba2007time} The FMO method in GAMESS\cite{schmidt1993general} utilizes a novel parallelization scheme\cite{fedorov2004new} to allow computations to be carried out efficiently on desktop computers as well as large scale super computers.\cite{fletcher2012large}Fragmentation can occur across covalent bonds using either the Hybrid Orbital Projection (HOP)\cite{nakano2000fragment} or Adapted Frozen Orbital (AFO)\cite{fedorov2008covalent,fedorov2009analytic} method. The EFMO method, also available in GAMESS, neglects the Coulomb bath from FMO and replaces it with classical terms to improve the computational speed. The input to EFMO and FMO are largely identical.

Often, the input files for the FMO method are more complex than the
regular \emph{ab initio} input files. The reason for this
complexity is that complete knowledge about the individual fragments of
interest are required, i.e. the atom indices that make up the fragment which might not be in any specific order in an input coordinate file, the integer fragment charges and level of theory. For a molecular cluster, each individual molecule can be considered a single fragment, for polymers a sub-unit of that polymer could make up a fragment whereas for proteins each individual residue can be considered a fragment. Fragmentation across covalent bonds adds more complexity: One must now also consider
chemically reasonable places of fragmentation (do not break conjugation, etc.) which itself requires manual inspection of the structure of interest. Different
systems have different complexities. For example, setting up a fragment calculation on a simple system consisting of three water molecules is feasible to prepare manually, but a protein with thousands of atoms is not. Consider also the case of
multiple layers which many of the methods support (see for example Ref~\citenum{fedorov2005multilayer}) in similar spirit to the ONIOM
method\cite{svensson1996oniom,dapprich1999new} where one (or several) lower
level layer(s) are used for some chemically irrelevant parts of a system but
their effect on a higher level layer, which is used for the chemically
interesting part, is needed. The assignment of fragments to individual layers
usually based on the distance to a point of interest, is also no minor task
when you have hundreds of fragments.

The need for automated tools which can setup calculations for a variety of
systems (proteins, molecular clusters, polysaccharides, etc.) and automating
the above tasks is thus of utmost importance if these methods are to become
viable tools that computational chemists use routinely in their research.

Software tools to prepare fragment method input files are already present but they differ greatly in their applicability and flexibility regarding different systems. The FMOutil\cite{fmoutil} package is supplied with the FMO method in GAMESS. It only supports fragmenting proteins/peptides and is dependent on a standardized PDB file format. It does include support for enabling multi-layered FMO calculations and letting the user choose whether to include solvation or not. Of more general applicability is the Facio\cite{suenaga2005facio, suenaga2008development} tool which supports the generation of FMO input files. The user can choose between fragmenting peptides, saccharides and nucleotides along with more specific options regarding the computational details such as level of theory and memory requirement. The main strength of Facio, however, is that it is a graphical user interface and one can define custom fragmentation bonds by using the mouse. While the FMOutil software is released under an open source license and can be run on any computer in a terminal, the Facio tool is closed source and available for Windows only. 

We present an open source fragmentation methodology, an accompanying command line tool and a corresponding web service called FragIt that enables one to easily fragment any molecule or system of interest using predefined (or custom) patterns to locate fragmentation points. As output, FragIt creates an input file to the FMO method in GAMESS with reasonable defaults so the calculation can be started directly. FragIt can be extended to write input files for other (fragment) methods and new patterns of fragmentation can be created and tested without changing the source code. The only requirement for the input is that the structure of interest is protonated correctly in advance according to the problem of interest which can be achieved by tools such as PDB2PQR\cite{dolinsky2004pdb2pqr,dolinsky2007pdb2pqr}. 

We have tested FragIt on several artificial and naturally occurring proteins with patterns of fragmentation to make reasonable fragments both in terms of the involved chemistry and size. We compare resulting fragmentation properties of FragIt and FMOutil and highlight similarities and differences. We also demonstrate that the fragmentation methodology is able to fragment a string of DNA and a polysaccharide successfully given the appropriate patterns. When SMARTS patterns are not possible, FragIt supports manual definition of fragmentation points. As an example, we fragment the Leucoemeraldine state of Polyaniline.

\section*{Design and Implementation}
The FragIt algorithm was initially inspired by the RECAP\cite{lewell1998recap} algorithm. The fragmentation algorithm (outlined in Figure~\ref{fig:figalgorithm} and
discussed below) will take any file format supported by Open Babel\cite{o2011open,openbabel} and fragment it, i.e. assign individual atoms to fragments, calculate the integer
fragment charge from partial atomic charges in a fragment, locate atoms which define the
boundaries of fragments and finally write an input file to the FMO method in
GAMESS. The details of the implementation of the fragmentation algorithm is described below. The only required user input is a properly protonated and
chemically reasonable structure.

To search for chemically reasonable places to fragment in molecules and to not rely on a single file format, we have based the fragmentation method on the SMiles ARbitrary Target Specification
(SMARTS)\cite{james2006daylight} language which enables us to make substructure
searches in molecules. For our specific needs, we have to find atomic species
which are written in SMARTS as \texttt{[M]}. Here, \texttt{M} is an atomic
primitive (an element such as a carbon atom). We are interested in locating
pairs of atoms connected by a covalent bond through which we wish to define
boundaries of fragments. A pair of atomic primitives is written as
\texttt{[M][N]} where \texttt{M} and \texttt{N} are both atomic primitives.
However, this is not nearly as flexible as we want since the chemical
environment for different bonds vary, so to build atomic primitive
environments we use the \texttt{\$()} operator to define
such environments as \texttt{[\$(MLHS)][\$(MRHS)]}. Here, \texttt{MLHS} and
\texttt{MRHS} are general SMARTS match patterns for the left hand side and
right hand side of a bond. The first atom in \texttt{MLHS} is covalently bound
to the first atom in \texttt{MRHS}. For instance, to match atoms on each side
of a peptide bond we might use \texttt{[\$(CC)][\$(NC)]} which would match a
Carbon connected to another Carbon on one side to a Nitrogen connected to a
Carbon on the other side (see Figure~\ref{fig:peptidebond}).

Due to the general application of SMARTS patterns, one can imagine protecting
certain parts of a molecule from fragmentation is useful, say for a ligand in a
protease or a specific residue in a protein. In FragIt, this protection is enforced via protection patterns which
are SMARTS patterns to match atomic primitives as above.

FragIt is implemented in Python\cite{python} and
relies heavily on the use of the Open Babel API and its ability to be accessed through a
SWIG\cite{swig} exposed Python interface\cite{openbabelpython}.

\subsection*{The Fragmentation Algorithm} The fragmentation algorithm is
outlined in Figure~\ref{fig:figalgorithm}. Initially, the structure of interest is loaded via
Open~Babel and the partial charges of all atoms are obtained from the MMFF94
force-field via the \texttt{OBChargeModel} class. The MMFF94 force-field is specifically chosen in this work since it contains atom types aimed at systems of biochemical interest. However, Open Babel does include other force-fields which are probably more suited to inorganic systems. After having obtained the charges, potential atoms which should be protected by the appropriate SMARTS patterns
(Table~\ref{tbl:fragpatterns}) are located. Hereafter follows the fragmentation procedure
which fragments the system according to the fragmentation SMARTS patterns or
explicitly defined valid pairs of atoms. In all cases SMARTS are handled by the
\texttt{OBSmartsPattern} class from the Open~Babel API and we obtain atom
indices of fragmentation points directly because we base the search on atomic primitives. To obtain the atoms that constitute a single
fragment, we use the atom indices of the fragmentation points found above and with the \texttt{FindChildren} method of the \texttt{OBMol} class to extract all atoms between the two.

The fragmentation algorithm supports grouping neighboring fragments together.
We have chosen an implementation which is a combination of two or more adjacent
fragments into one. This combination of fragments has one benefit from a
computational point of view, and that is to increase the accuracy of the
computation.

Once fragments are identified (and optionally grouped), we can assign fragment
charges by counting the partial atomic charges in each fragment we obtained earlier.

Finally we write the input file for the FMO method in GAMESS.

\subsection*{Writing the Input Files} When the fragmentation data has been
obtained an input file is written to disk for the FMO method in GAMESS. There
are several options available when writing the input file. First,
multi-layered input file generation is supported by selecting a
fragment which is the \emph{central fragment}. From this fragment the
assignment to layers are calculated using the \emph{minimum} distance $R_{IJ}$ from fragment $I$, where $I$ is the central fragment, to
all other fragments $J$ based on a user defined distance that separates one
layer from another.

In addition to the above, support for the FMO Frozen Density (FMO/FD)\cite{fedorov2011geometry} method is also included. In FMO/FD one has an active region in which we wish to do a geometry optimization, a buffer region to help relax the density of the active region and a frozen region where the density is kept frozen after initial convergence. One defines a central fragment as above and a distance. Fragments which have atoms within
this distance from the central fragment should all be enabled for geometry
optimization. Around this active region, a buffer zone is constructed similarly
with a new distance within which atoms (and fragments) are considered a buffer.
The rest is frozen.

The input file can be run directly in GAMESS because we provide sensible defaults which are not FMO specific (an energy calculation using RHF with a 3-21G basis set, 1 GB of memory per core, etc.). Specifically for FMO, only the AFO bonding scheme is supported since the original HOP formulation requires manual generation of molecular orbitals in the bonding region if one chooses exotic places to fragment. The HOP scheme is also dependent on an extra basis set input which is automatically done with AFO. While the defaults are only meant to get novice EFMO and FMO users started using fragment based methods we note that manual changes to the input files for a production run is needed.


Finally, the ability to write an extra PyMOL\cite{pymol} and/or Jmol\cite{jmol} script to visually inspect the fragmentation (and optionally layering) is included. The scripts are based on templates which contains the markup needed to make the programs visually display the fragmentation. The fragment information from FragIt is then used to group atoms into fragments for visual display.

Figures in this paper were generated using the PyMOL functionality. The web service uses the Jmol functionality.

\subsection*{Description of the Test Dataset}
To illustrate the applicability of FragIt to different molecules using
different patterns, we show results for a molecular cluster consisting of water
and a solute, Chignolin (PDB: 1UAO), Tryptophan-cage (PDB: 1L2Y), Human
Parathyroid Hormone (HPH) residues 1-34 (PDB: 1ET1A), Crambine (PDB: 1CRN),
GluR2 ligand binding core (PDB: 1FTJ) as well as several neutral methyl-capped
$\alpha$-helix and $\beta$-sheet alanine structures from Fedorov \emph{et
al.}\cite{fedorov2009fragment} and finally a $\beta$-cyclodextrin (extracted
from PDB: 3CGT). The neutral methyl-capped $\alpha$-helices and $\beta$-sheets
are good to show FragIt's capability to fragment and group fragments together.
Chignolin, the Tryptophan-cage, Crambine and Glur2 are real world
proteins and have charged termini. These will help show that FragIt can protect
various parts of a protein (in this case NH3$^{+}$ groups) by grouping them
with nearby fragments to increase the computational accuracy. Crambine and
Glur2 also include Sulfur-bridges. The structure of B-DNA was obtain from Georgia State University\cite{dnafile}. Leucoemeraldine was built in Avogadro\cite{avogadro}. Lastly, the polysaccharide
$\beta$-cyclodextrin is included to illustrate that by simply including the
appropriate patterns of fragmentation these are fragmented as well. These files are available in Files~S1.

The crystal structures of the proteins and $\beta$-cyclodextrin were
protonated using PDB2PQR\cite{dolinsky2004pdb2pqr,dolinsky2007pdb2pqr} at
pH$=7$.

\section*{Results}
We now show four use cases of FragIt. First, we use it extensively to fragment proteins where we show many combinations of properties and compare with the FMOutil program. We then show that FragIt fragments polysaccharides, strands of DNA and other polymers as well. For an in-depth discussion of fragmentation patterns we refer to the Design and Implementation section. Groups of atoms not connected by a covalent bond are automatically grouped into separate fragments.  So for example, any water molecules are treated as individual fragments.

\subsection*{Proteins}
The fragmentation pattern used for peptide bonds is shown in Table~\ref{tbl:fragpatterns} and illustrated in Figure~\ref{fig:figprotpat}[c-e]. Earlier work has shown\cite{nakano2000fragment,nakano2002fragment} that fragmentation at the C$^{\alpha}$-C' bond leads to higher accuracy energies for the FMO method compared to fragmentation at the semi-conjugated peptide bond. One has to also take care that fragments will not be too small and cause unphysical behavior. To facilitate this, we include several patterns of protection in FragIt. A protection pattern locates parts of a structure which must not be fragmented. The fragmentation pattern will generate small fragments at the N-termini (Figure~\ref{fig:figprotpat}c) but this fragment is chemically too small because of the positive charge right next to a fragmentation point which will result in a too large inter-fragment charge-transfer. Furthermore, because of the way fragment boundaries are made using AFO uneven charges arising from the N-termini are also taken care of using protection patterns. The protection patterns (also listed in Table~\ref{tbl:fragpatterns}) will protect both a charged and a neutral N terminus. Protection of the C-terminus is implicitly built into the fragmentation pattern.

Shown in Table~\ref{tbl:fragmentationproteins} are results for the several proteins. We list the number of residues for each protein as well as the number of fragments obtained after fragmentation. For the $\alpha$-helices and $\beta$-sheets which are capped with methyl groups, we obtain as many fragments as residues. This is different from Chignolin, Tryptophan-cage and HPH where the protection patterns match the N-termini and (for one residue per fragment) make one less fragment than the number of residues. For GluR2, the results are similar, but there are also seven water molecules as well as two chains (which both get protected) yielding a total of 257 fragments in the protein plus 7 water fragments totaling 264. The column $N_\mathrm{A}^\mathrm{max}$ in Table~\ref{tbl:fragmentationproteins} lists the maximum number of atoms in any fragment. For Crambine, the number of residues is 46 and the resulting number of fragments (for one residue per fragment) is 42 because one is protected and the three disulfide bridges are combined into three (rather than six) fragments. Disulfide bonds are not subject to automatic fragmentation in FragIt unless a specific pattern is supplied. This is an intentional side-effect of using fragment based patterns because the S-S bond is comparatively very polarizable and therefore a poor choice for a covalent link.

Larger fragments leads to more accurate FMO results (see for example Ref~\citenum{fedorov2004importance} for more information) and FragIt allows grouping of fragments where two covalently bonded fragments (such as two adjacent amino acids) can be grouped in to one fragment. This is illustrated in Figure~\ref{fig:chignolinoptions} where fragments are colored to distinguish them. Because of using only 6 different colors, some fragments (although different) will have the same color. Here, a) shows Chignolin is fragmented without protection patterns using one residue per fragment. This results in 10 fragments with fragment sizes between 7 and 24 atoms (see Table~\ref{tbl:fragoption}) which is the same as the default behavior of FMOutil. b) shows the use of a protection pattern to prevent the first peptide bond from being cut. This is the default setting and gives 9 fragments with fragment sizes from 7 atoms to 28 atoms. We do note that the default setting gives slightly more imbalanced fragment sizes when thinking about parallelization strategy. To further improve fragment size balance one can choose to merge all Glycine residues with preceding fragments. This is a reasonable strategy to improve accuracy without added computational cost and is also possible in the FMOutil. With the default fragmentation behavior of FragIt and merging of Glycine residues the result is 6 fragments with fragment sizes ranging from 14 atoms to 28 atoms whereas not including protection patterns but merging gives rise to 7 fragments with fragment sizes from 7 atoms to 34 atoms. Correspondingly, merging Glycine residues with FMOutil yields 7 fragments with sizes 12 atoms to 34 atoms. The difference between FragIt and FMOutil lies in the way they see Glycine. FMOutil uses knowledge from residues to merge Glycine whereas FragIt relies on a pattern to find it. After fragmentation, the SMARTS pattern used does not recognize the N-terminal in Chignolin as a Glycine which gives rise to the discrepancy.

Including grouping but neglecting the protection pattern and merging gives rise to 5 fragments with a maximum of 37 atoms in a fragment (with the smallest fragment having 21 atoms). This is shown in Figure~\ref{fig:chignolinoptions}c) and the same fragment sizes that FMOutil does (Table~\ref{tbl:fragoption}). Compare this to d) where the combination of protection patterns and grouping leads to large size differences in fragments (minimum fragment size is 10 atoms and the maximum is 40 atoms). Different combinations of options can lead to very different fragmentation possibilities and care should be taken to fragment a system in the most sensible way in which the physics of the individual fragments is properly described.

\subsection*{Polysaccharides}
Table~\ref{tbl:fragpatterns} lists the fragmentation pattern we have included to fragment chains of D-galactopyranose. This pattern is specifically aimed at fragmenting cyclodextrins which are common in the design of artificial enzymes. The pattern explicitly matches the $\ensuremath{\mathrm{CH_2OH}}$ side chain. Figure~\ref{fig:figsachpat}[a] shows a trimer of D-galactopyranose and Figures~\ref{fig:figsachpat}[b-d] shows how the trimer is fragmented. The pattern takes an $N$-mer of sugar and converts it to $N$ fragments, see Table~\ref{tbl:fragmentationproteins}. This also works for the cyclodextrins without any modifications. We did not observe any fragments during our tests that were of such size that protection patterns were necessary.

To fragment other sugars, one could modify the existing pattern to suit ones needs.

\subsection*{DNA}
Although proteins has been the primary target of the majority of the fragment based methods, another interesting system to investigate is DNA. The nucleotides of DNA are connected to the DNA backbone which is made from sugars and phosphate groups connected by covalent bonds. We include a basic pattern (see Table~\ref{tbl:fragpatterns}) to fragment the backbone of DNA which can be observed in Figure~\ref{fig:dna}. The 24 nucleotides were fragmented into 26 individual fragments and except from the two 9 atom fragments which make up the termini of each backbone (and could be protected by additional protection patterns) fragment sizes are 27 to 33 atoms depending on the type of nucleotide involved.

\subsection*{Other Polymers}
Under some circumstances, a fragmentation pattern cannot be supplied, for instance when looking at Polyaniline since the right hand side and the left hand side of bond are the same which leads to multiple unwanted matches and very small fragments. To fragment such systems, it is possible to manually define pairs of atoms between which there should be a fragmented bond (see below). To show this feature, we have fragmented Leucoemeraldine, the fully reduced state of a chain of Alinine monomers (shown in Figure~\ref{fig:leuc} and included in the supporting information, Files S1). This makes it possible to fragment systems even without having developed a fragmentation pattern beforehand. This option can be combined with regular fragmentation patterns allowing for very specialized fragmentation setups.

Another option which we might consider in the future is to implement a minimum fragment size option which could allow for these unwanted matches

\subsection*{Installation and Usage}
The FragIt software is installed by downloading the source code from the project homepage (see below). FragIt requires Python v2.4 or later (not 3.X), Numpy\cite{numpy} v1.5 (or later) and the Open Babel framework compiled with Python bindings version 2.3 (or later). Once these requirements are satisfied, FragIt can be executed from the command line. Options which are described in detail above is easily accessed through arguments to the command line executable. The web version requires Java to run Jmol, but otherwise no installation on a client computer.

To use FragIt invoke the executable and supply the structure which we wish to fragment (here Chignolin)
\begin{verbatim}
fragit 1uao.pdb
\end{verbatim}
to generate an input file using the default settings (illustrated in Figure~\ref{fig:chignolinoptions}b). The default settings include fragmentation patterns for peptide bonds and sugar bonds as well as the listed protection patterns. To disable the use protection patterns use the \texttt{--disable-protection} command line option. All available options can be changed via command line options to the FragIt executable or via configuration files. To generate a configuration file named \texttt{my.conf} with the default fragmentation settings use the \texttt{make-config} option on a molecule of interest
\begin{verbatim}
fragit --make-config=my.conf
\end{verbatim}
The contents of this file has all modifiable options (see supporting information) and can be supplied to FragIt via the \texttt{use-config} option. To test new patterns for instance, this is the preferred approach since no tinkering with the FragIt source code is needed.

To invoke layered input files as described above, one must specify a central fragment (\texttt{--output-central-fragment}) as well as a distance (given by \texttt{--output-boundaries}) within which all fragments will be promoted to a higher level layer. This is accomplished by
\begin{verbatim}
fragit --output-central-fragment=1 --output-boundaries=3.0 1uao.pdb
\end{verbatim}
which takes fragments within 3 {\AA} of the first fragment and promotes it to a higher level layer.

To activate active, buffer and frozen regions, one must again specify both a central fragment and boundaries for higher level layers which are allowed to move\cite{fedorov2011geometry}. Moreover, another distance within which atoms (and their fragments) are considered active is specified by the \texttt{--output-active-distance} option. Lastly, a buffer region of fragments between the active and the frozen is defined by a final distance ({\texttt{--output-buffer-distance})
\begin{verbatim}
fragit --output-central-fragment=1 --output-boundaries=3.0 \
       --output-active-distance=2.0 --output-buffer-distance=3.0 1uao.pdb
\end{verbatim}
which first will generate layers based on the boundary settings as above, then find active fragments within 2.0 {\AA} from the central fragment, and finally create a buffer region around the active fragments within 3.0 {\AA} from any active fragment. In this final step, some previously lower level layers may be promoted to higher level ones. The result is shown in Figure~\ref{fig:activebuffer}b. By changing the \texttt{--output-active-distance} option to 1.0, one obtains Figure~\ref{fig:activebuffer}a.

To group $N$ consecutive fragments, one can use the \texttt{--g N} option. Usually, $N$ would be 2 but FragIt supports any positive integer. To only merge Glycine the \texttt{--merge-glycine} option is available.

Lastly, to manually specify points of fragmentation one has to use configuration files and define the \texttt{pairs} string in \texttt{explicitfragmentpairs} group. The format is \texttt{pairs = A,B;C,D;} where \texttt{A}, \texttt{B}, \texttt{C} and \texttt{D} are all atom indices and \texttt{A,B} is an atom pair between one wishes that there is an explicit bond.


\section*{Availability and Future Directions}
The web service\cite{fragitserver} available at \texttt{www.fragit.org} enables users to upload their structure, fragment it and download the resulting input file to GAMESS. The user is able to visually inspect (see Figure~\ref{fig:webservice}) the fragmentation using Jmol and make simple changes such as using multiple residues per fragment or enable layered input or optimization as discussed above. We are actively implementing features from the command line executable to work with the web service.

For greater flexibility, we strongly encourage the use of the command line tool which can be downloaded from the above URL or accessed from the development source at \texttt{www.github.com/FragIt/}.

The FragIt source code is distributed under an open source license (GPL, version 2 or later) and users of the FragIt code are encouraged to submit changes and additions, especially for their own (fragmentation) methods. It would also be possible to combine FragIt with other open source graphical tools such as Avogadro or even PyMOL, providing an alternative to the Facio software.

We plan on using and extending FragIt in the future with new patterns of fragmentation and methods as our research heads in new directions.

\section*{Acknowledgments}
CS kindly thanks Anders Steen Christensen and Luca De Vico for helpful comments and
suggestions during the writing of this paper. Luca De Vico beta tested
FragIt extensively. CS also kindly thank Jacob Poehlsgaard from the Department
of Pharmaceutical Sciences, University of Copenhagen for testing out FragIt
early on.

\bibliography{article_plos}

\section*{Figure Legends}
\newpage
\begin{figure}[!ht]
\begin{center}
\includegraphics[width=8cm]{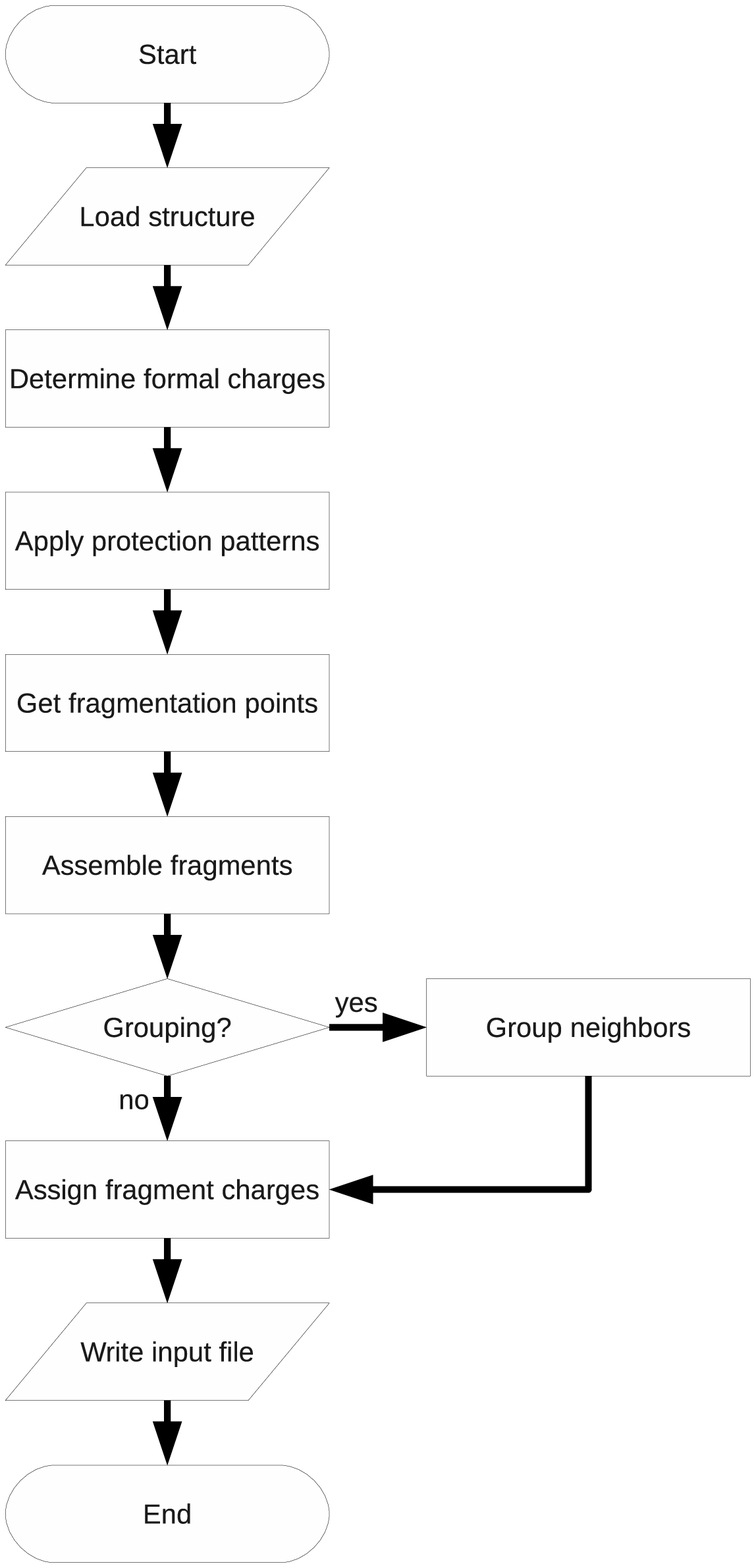}
\end{center}
\caption{
{\bf Scheme of the FragIt Algorithm.}}
\label{fig:figalgorithm}
\end{figure}

\newpage
\begin{figure}[!ht]
\begin{center}
\includegraphics[width=8cm]{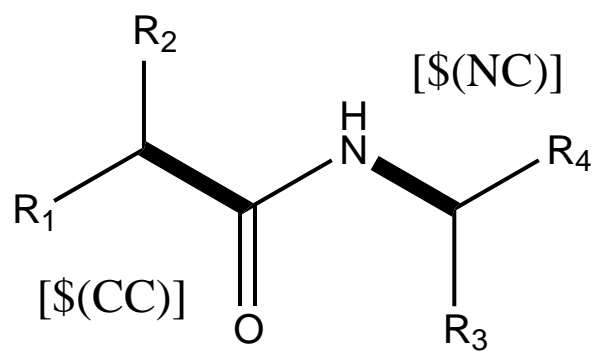}
\end{center}
\caption{
{\bf Pattern matching on a peptide bond using SMARTS}.
}
\label{fig:peptidebond}
\end{figure}

\newpage
\begin{figure}[!ht]
\begin{center}
\includegraphics[width=8cm]{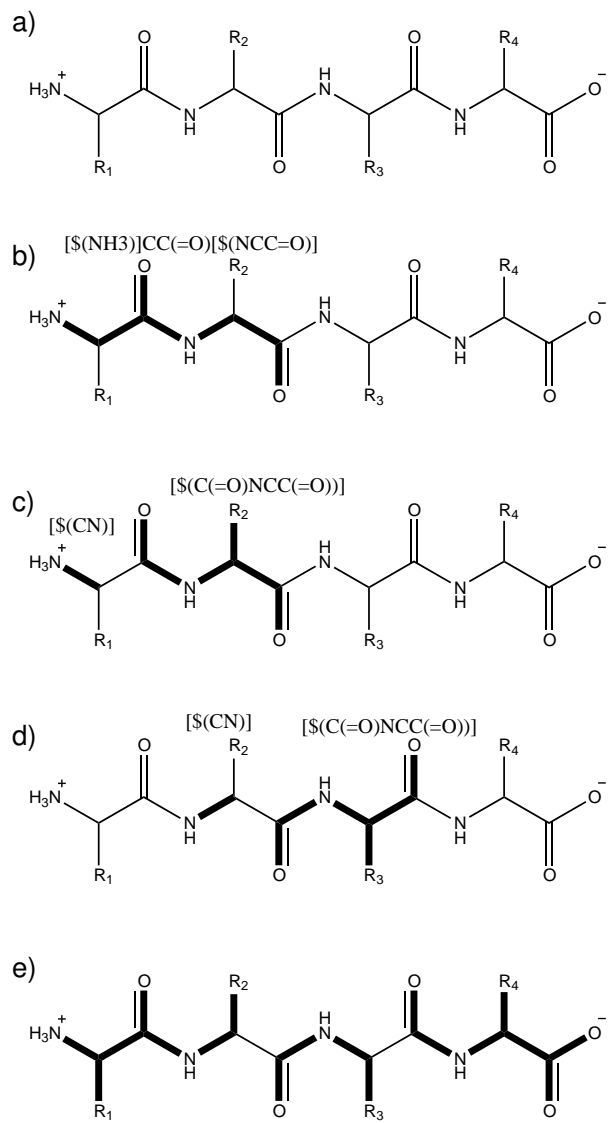}
\end{center}
\caption{
{\bf Protein Backbone Fragmentation Example.} An example of how fragmentation and protection is carried out using SMARTS patterns on a protein in FragIt. Illustration a) is the uncorrupted protein backbone with side-chains R$_1$ through R$_4$, b) shows how a protection pattern matches atoms very specifically. c) and d) shows examples of fragmentation using the standard peptide pattern supplied in this work. Finally, e) is the final fragmentation when all fragmentation (4 fragments) and protection (1 fragment) is carried out, resulting in 3 fragments.
}
\label{fig:figprotpat}
\end{figure}

\newpage
\begin{figure}[!ht]
\begin{center}
\includegraphics[width=17cm]{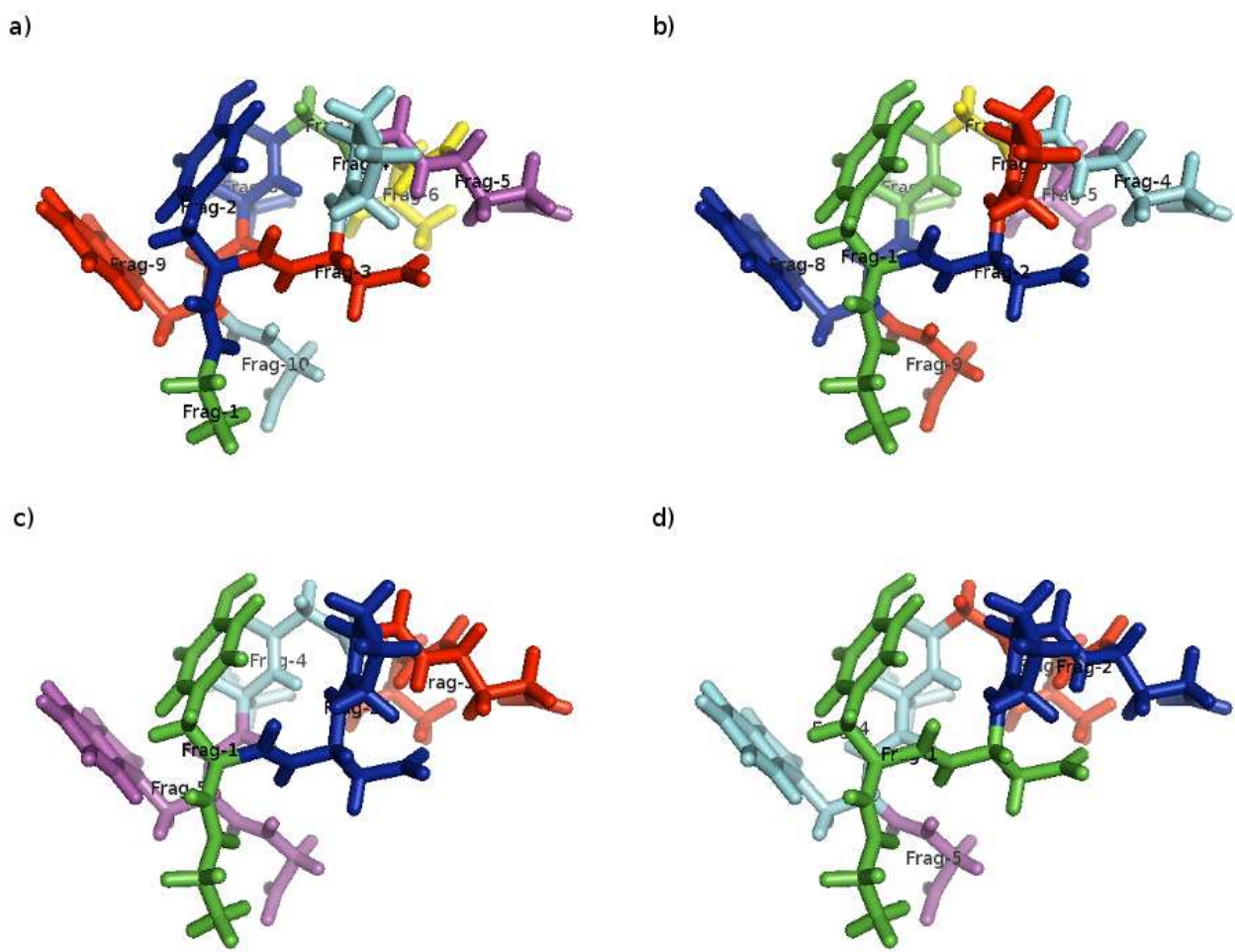}
\end{center}
\caption{
{\bf Different fragmentation options illustrated for Chignolin.} Here shown a) without protection and no grouping, b) with protection and no grouping, c) without protection but in groups of two residues per fragment and d) with protection and using two residues per fragment. We use a six color coloring scheme resulting in different fragments may have the same color.
}
\label{fig:chignolinoptions}
\end{figure}

\newpage
\begin{figure}[!ht]
\begin{center}
\includegraphics[width=8cm]{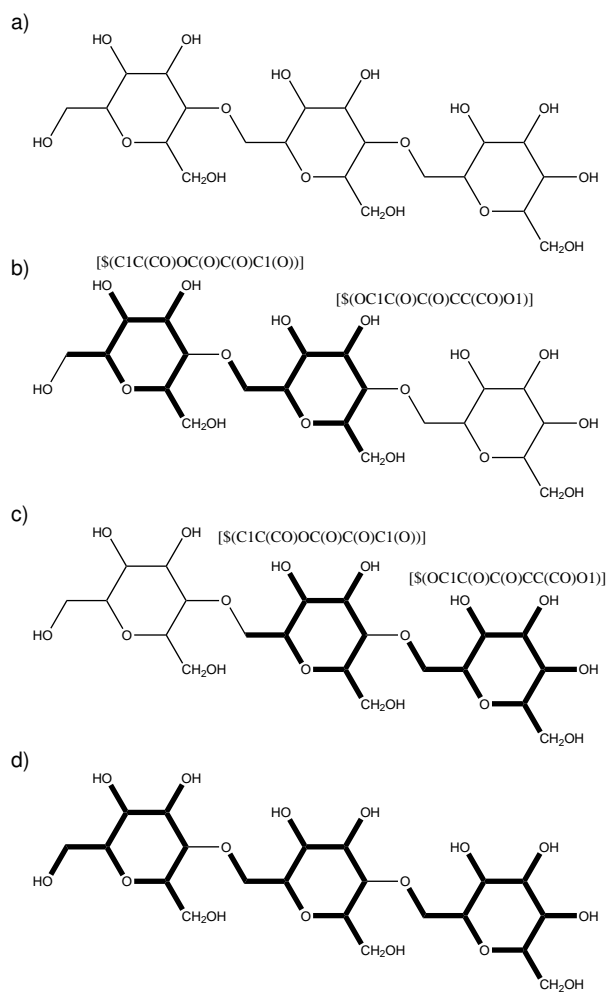}
\end{center}
\caption{
{\bf Sugar Fragmentation Example.} An example of how fragmentation is carried using SMARTS patterns on a polysaccharide molecule in FragIt. Illustration a) is the polysaccharide molecule of interest, in this case an D-galactopyranose trimer. b) and c) shows examples of fragmentation using the polysaccharide fragmentation pattern in this work. Finally, d) shows the final fragmentation of 3 fragments.
}
\label{fig:figsachpat}
\end{figure}

\newpage
\begin{figure}[!ht]
\begin{center}
\includegraphics[width=8cm]{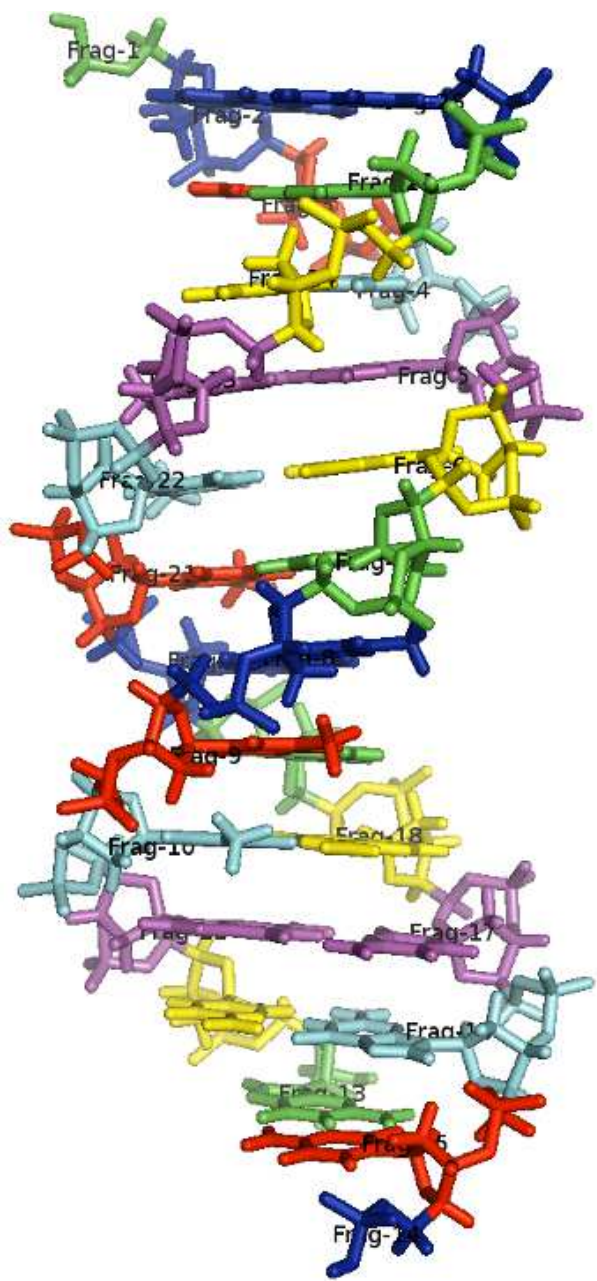}
\end{center}
\caption{
{\bf Fragmentation of DNA using FragIt.} We use a six color coloring scheme resulting in different fragments may have the same color.}
\label{fig:dna}
\end{figure}

\newpage
\begin{figure}[!ht]
\begin{center}
\includegraphics[width=17cm]{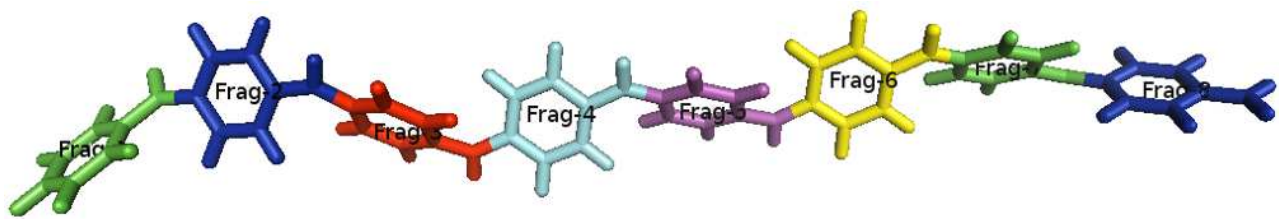}
\end{center}
\caption{
{\bf Fragmentation of Leucoemeraldine (a Polyaniline) using FragIt.} This particular example uses no fragmentation patterns but rather explicitly defined points of fragmentation by the end user.}
\label{fig:leuc}
\end{figure}

\newpage
\begin{figure}[!ht]
\begin{center}
\includegraphics[width=17cm]{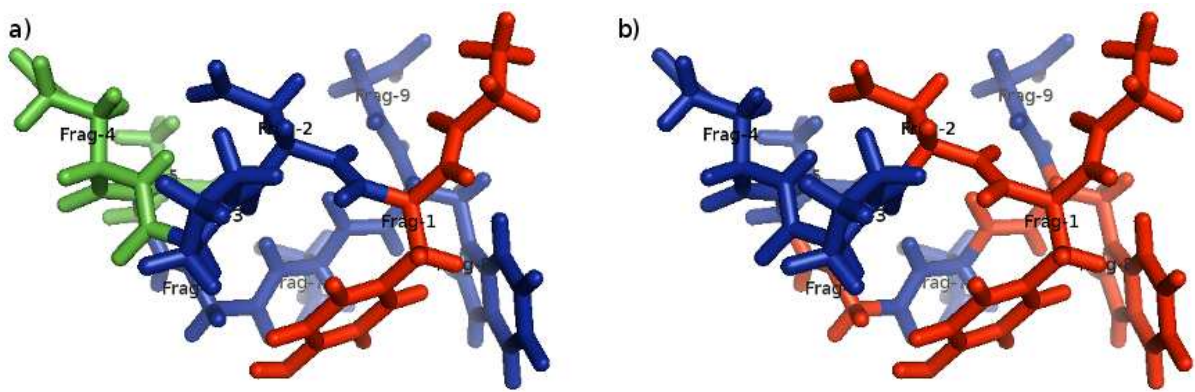}
\end{center}
\caption{
{\bf Different active, inactive and frozen regions in Chignolin}. Regions are color-coded according to their function: red is active, blue is buffer and green is frozen. In a) fragment 1 is in the active region, fragments 2, 3, 6-9 are buffer region fragments and fragments 4 and 5 are in the frozen region. In b), fragments 1,2,6 and 8 are active fragments while fragments 3,4,5,7 and 9 are buffer region fragments. There are is no frozen region in b).
}
\label{fig:activebuffer}
\end{figure}


\newpage
\begin{figure}[!ht]
\begin{center}
\includegraphics[width=17cm]{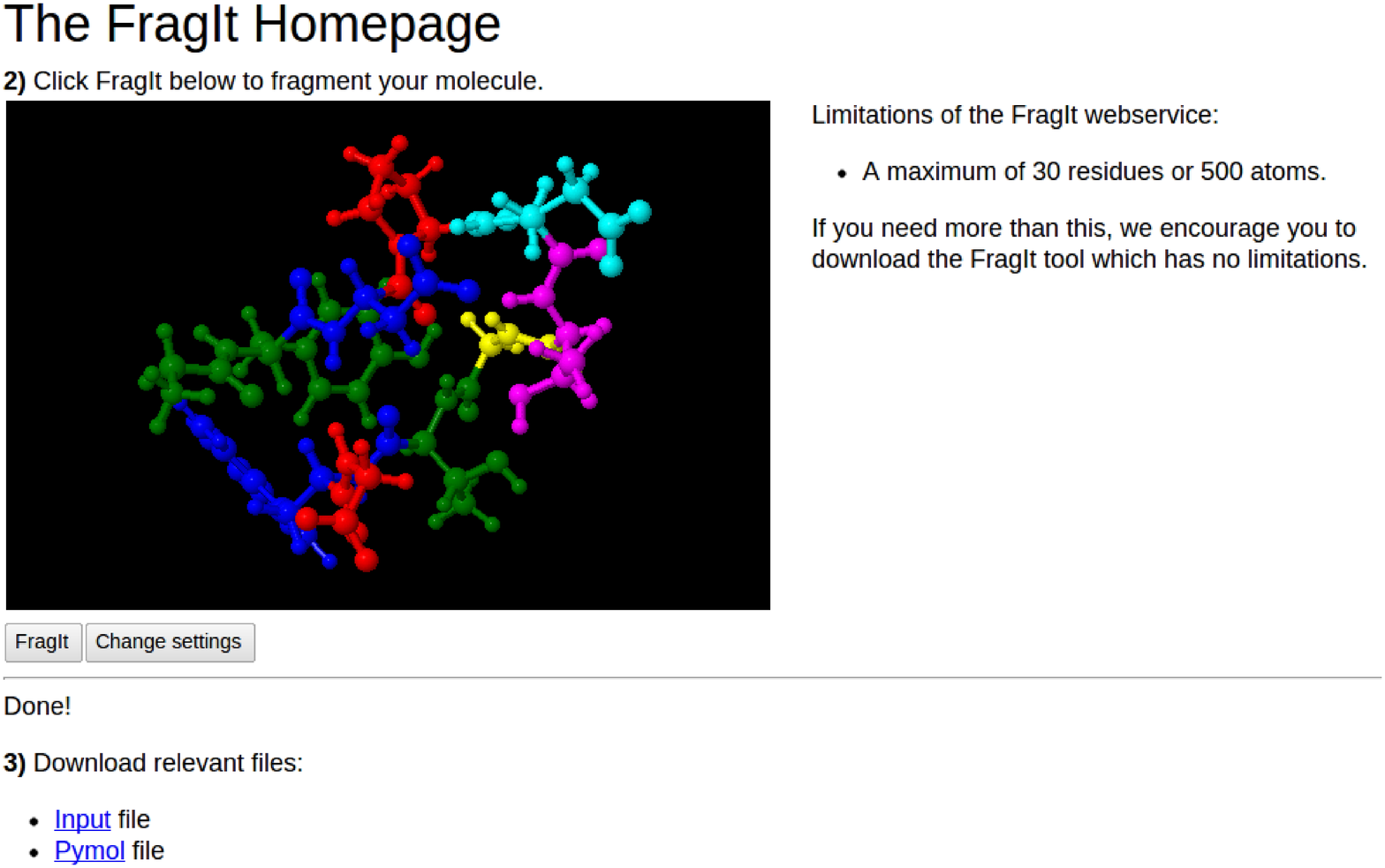}
\end{center}
\caption{
{\bf Fragmentation of Chignolin using the web service}.
}
\label{fig:webservice}
\end{figure}

\section*{Tables}
\begin{table}[!ht]
\caption{
\bf{Patterns of Fragmentation}}
\begin{tabular}{cccll}
\hline & Description & Match & LHS & RHS \\ \hline

 Protein & fragmentation & -C$^\alpha$\textbar-C'- & [\$(CN)] & [\$(C(=O)NCC(=O))] \\
 & NH$_2$ protection & -NH$_2$ & [\$(NH2)] & CC(=O)[\$(NCC=O)] \\
 & NH$_3^+$ protection & -NH$_3^+$ & [\$(NH3)] & CC(=O)[\$(NCC=O)] \\ \hline
 Sugar & fragmentation & -C\textbar-C- & [\$(C1C(CO)OC(O)C(O)C1(O))] & [\$(OC1C(O)C(O)CC(CO)O1)] \\ \hline
 B-DNA & fragmentation & -C\textbar-C- & [\$(CCOP)][\$(CC1OCCC1)] \\
\end{tabular}
\begin{flushleft}Patterns used in FragIt for different types of chemical systems. The patterns for proteins include protection patterns to increase accuracy by eliminating very small fragments and the default fragmentation pattern makes sure to keep the quasi-conjugated nature of the peptide bond intact. For polysaccharides a single pattern to match $\alpha$-D-galactopyranose units, the subunits of cyclodextrins, is included. We also include a pattern to fragment the backbone of DNA.
\end{flushleft}
\label{tbl:fragpatterns}
\end{table}

\newpage
\begin{table}[!ht]
\caption{
\bf{Default Fragmentation Results}}
\begin{tabular}{lccccc|l}
Protein & $N_\mathrm{U}$ & $N_\mathrm{frag} (1/2)$ & $N^\mathrm{max}_\mathrm{A}(1/2)$ & $Q$ & $N_\mathrm{S-S}$ & Comments \\ \hline
 $\alpha$-(ALA)$_{10}$ & 10  & 10/5    & 18/28&  0 & 0 & capped with methylene\\
 $\alpha$-(ALA)$_{20}$ & 20  & 20/10   & 18/28&  0 & 0 & capped with methylene\\
 $\alpha$-(ALA)$_{40}$ & 40  & 40/20   & 18/28&  0 & 0 & capped with methylene\\
 $\beta$-(ALA)$_{10}$  & 10  & 10/5    & 18/28&  0 & 0 & capped with methylene\\
 $\beta$-(ALA)$_{20}$  & 20  & 20/10   & 18/28&  0 & 0 & capped with methylene\\
 $\beta$-(ALA)$_{40}$  & 40  & 40/20   & 18/28&  0 & 0 & capped with methylene\\
 Chignolin             & 10  & 9/5     & 28/34&  -2 & 0 & \\
 Tryptophan-cage       & 20  & 19/10   & 33/41&  +1 & 0 & \\
 HPH                   & 34  & 33/17   & 26/46&  +1 & 0 & \\
 Crambine              & 46  & 42/23   & 28/43&  0 & 3 & \\
 Glur2                 & 259 & 264/138 & 34/46&  +4 & 1 & {\small 7 waters included} \\
 $\beta$-cyclodextrin  & 7   & 7/4     & 21/42&  0 & 0 & \\
 B-DNA                 & 24  & 26/14   & 33/64 &  0 & 0 & \\
 Leucoemeraldine       & 8   & 8/4     & 13/25 &  0 & 0 & \\ \hline
\end{tabular}
\begin{flushleft}Fragmentation results for some selected proteins, enzymes and sugar molecules. We show the number of units $N_\mathrm{U}$ in the molecules of interest, i.e. residues for proteins or sugars in the cyclodextrins, the number of resulting fragments $N_\mathrm{frag}$ both without (1) and with (2) grouping with the neighbor, the maximum number of atoms in a fragment both without (1) and with (2) grouping with the neighbor, the overall charge $Q$ of the system at pH=7 and the number of sulfur bridges $N_\mathrm{S-S}$ which in FragIt is automatically treated as one fragment.
\end{flushleft}
\label{tbl:fragmentationproteins}
\end{table}

}
\newpage
\begin{table}[!ht]
\caption{
\bf{Fragmentation of Chignolin using various options for FragIt and FMOutil}}
\begin{tabular}{rrr|l}
\multicolumn{4}{c}{FragIt} \\ \hline
$N_\mathrm{frag}$ & $N_A^\mathrm{min}$ & $N_A^\mathrm{max}$ & comment \\ \hline
9  & 7  & 28 & default \\
10 & 7  & 24 & no protection \\
6  & 14 & 28 & default + merge \\
7  & 7  & 34 & no protection + merge \\
5  & 10 & 40 & group in pairs \\
5  & 21 & 34 & group in pairs + no protection \\ \hline
\multicolumn{4}{c}{FMOutil} \\ \hline
10 & 7  & 24 & default \\
7  & 12 & 34 & default + merge \\
5  & 21 & 34 & group
\end{tabular}
\begin{flushleft}
\end{flushleft}
\label{tbl:fragoption}
\end{table}

\section*{Supporting Information Legends}
\begin{flushleft}
\textbf{Files S1.} Example files for FragIt. Included are the protein and protein-like structures and cyclodextrin discussed in the text. We also include examples of configuration files with and without protection patterns.
\end{flushleft}
\end{document}